\documentclass[aps,prd,onecolumn,superscriptaddress,nofootinbib,10pt,floatfix]{revtex4-2}
\usepackage{graphicx}
\usepackage{amsmath,amssymb,bm}
\usepackage{xcolor}
\definecolor{bibteal}{rgb}{0.0,0.5,0.5}
\newcommand{\dd}{\mathrm{d}}

\newcommand{\ts}{\tau_1^{*}}
\newtheorem{proposition}{Proposition}
\usepackage{hyperref}
\begin{document}

\title{Frenet--Serret invariants of timelike worldlines in Kerr spacetime:\texorpdfstring{\\}{ }
chart independence, horizon crossing and null-endpoint limits}

\author{Paulami Majumder}
\affiliation{Centre of Excellence in Space Sciences India, Indian Institute of Science
Education and Research Kolkata, Mohanpur, West Bengal 741246, India}
\email{paulami.majumder93@gmail.com}

\date{\today}

\begin{abstract}
We examine the Frenet--Serret (FS) curvature $\kappa$ and first torsion $\tau_1$ of timelike worldlines in Kerr spacetime, focusing mainly on non-Killing equatorial trajectories. Since these quantities are scalars attached to a worldline, changing coordinates cannot remove a divergence for that same curve. We demonstrate this by transforming one worldline between five coordinate charts and checking that its invariants agree, including in an off-equatorial test. We also show why simply reusing the same coordinate components for the tangent in two charts describes different curves.

For a non-unit tangent $v$, with norm $S=g(v,v)$, we derive the identity $\kappa^2=Q^2/S^3-q\!\cdot\!q/S^2$, where $q=\nabla_vv$ and $Q=q\!\cdot\!v$. When $Q$ tends to a nonzero limit and $q\!\cdot\!q$ remains finite, this gives $\kappa\sim |Q_0|S^{-3/2}$ as the curve approaches a null endpoint. Under the additional condition $S'\ne0$, we obtain a closed expression for the limiting value of $\tau_1$. We check these results using exact symbolic calculations for two Kerr--Schild spirals, including one with $a>M$, for which no event horizon exists. For Killing observers, we derive closed forms for $\kappa$ and $\tau_1$ and show that both become singular at light surfaces. Finally, we distinguish the FS torsion, which measures rotation relative to the acceleration direction, from gyroscope rotation rates defined using regular reference frames. For a geodesic plunge, the rate relative to the principal-null frame remains finite across the horizon, while the rate relative to the Carter-plane axis agrees with Marck’s precession rate. Our analysis is restricted mainly to equatorial examples; a numerical comparison in double-null coordinates is left for future work.
\end{abstract}

\maketitle


\section{Introduction}
\label{sec:intro}

Gyroscopic precession provides a way to study how spacetime curvature and rotation affect the motion of a body \cite{Schiff,GPB}. The Frenet--Serret (FS) formalism describes the geometry of a worldline through quantities such as its curvature $\kappa$ and first torsion $\tau_1$ \cite{IV,Sem97,NV,Sem99}. The curvature measures the bending of the trajectory, while the first torsion describes the rotation of the acceleration-adapted FS frame along the worldline. These quantities provide a geometrical description of motion and are independent of the coordinate system used to represent a given trajectory.

In Kerr spacetime, FS quantities have been studied mainly for circular motion and other trajectories associated with spacetime symmetries, known as Killing trajectories~\cite{Bini16,Sem98}. More general motion arises in an extreme-mass-ratio inspiral (EMRI), where a compact object orbits a much more massive black hole along an evolving trajectory that is generally neither exactly circular nor generated by a Killing symmetry \cite{BC,Dest,Piovano,Skoupy}. Describing gyroscopic precession along such trajectories therefore requires an analysis that goes beyond the special cases associated with spacetime symmetries.

In our earlier work, we developed the FS description for arbitrary timelike trajectories and studied gyroscopic precession near the horizons of Schwarzschild and Reissner--Nordstr\"om black holes \cite{MN23,MN25}. The present paper extends this investigation to Kerr spacetime. We focus on three related questions: how the FS invariants behave under coordinate transformations, how they approach limiting values when a timelike trajectory becomes null, and how the FS description relates to gyroscope rotation measured relative to a regular geometrical frame. We do not claim as new either the general treatment of non-Killing trajectories or the possibility of finite gyroscopic precession at a regular event horizon.

The first issue is to distinguish a change of coordinates from a change of trajectory. The FS curvature $\kappa$ and first torsion $\tau_1$ are scalars associated with a worldline. Their values must therefore agree when the same trajectory is described consistently in different coordinate systems. This requires transforming the tangent vector and the metric together. By contrast, imposing the same coordinate components for the tangent independently in Boyer--Lindquist (BL) and horizon-penetrating coordinates will generally define different worldlines. Differences in their FS invariants then reflect differences in the trajectories rather than a failure of coordinate invariance. We test this distinction by comparing the invariants in several coordinate systems, including an off-equatorial example (Sec.~\ref{sec:charts}).

We next study the behaviour of the FS invariants when a timelike trajectory approaches a null endpoint. To analyse this limit, we use a tangent vector $v^\mu$ that need not be normalised to unit proper time. Its squared norm is
\begin{equation}
    S \equiv g_{\mu\nu}v^\mu v^\nu,
    \label{eq:intro_S}
\end{equation}
where $g_{\mu\nu}$ is the spacetime metric. We use the metric signature $(+,-,-,-)$, so the trajectory is timelike when $S>0$ and approaches a null direction when $S\to0^+$. We also define the covariant acceleration associated with the curve parameter $s$ as
\begin{equation}
    q^\mu \equiv v^\nu\nabla_\nu v^\mu,
    \qquad
    Q \equiv q_\mu v^\mu
      = \frac{1}{2}\frac{dS}{ds},
    \label{eq:intro_qQ}
\end{equation}
where $\nabla_\nu$ denotes the covariant derivative. The scalar $Q$ measures the rate of change of the tangent norm along the curve.

These definitions lead to an exact identity for the FS curvature, given in Eq.~\eqref{eq:kappaS}. If $Q$ approaches a nonzero limit $Q_0$ and $q_\mu q^\mu$ remains finite, the curvature behaves as
\begin{equation}
    \kappa \sim |Q_0|S^{-3/2}
    \qquad \text{as} \qquad S\to0^+.
\end{equation}
This scaling is conditional and does not apply to every trajectory approaching a null direction. Under the additional assumption that $S'\neq0$ at the endpoint, where the prime denotes differentiation with respect to $s$, we also derive a closed expression for the limiting value of the first torsion (Proposition~\ref{prop:tau}). We test these results using two Kerr--Schild spiral trajectories and compare the predicted torsion limits with numerical calculations in seven additional cases. One of the two Kerr--Schild examples lies in the super-extremal regime $a>M$, where no event horizon exists. This allows us to investigate null-endpoint behaviour independently of the presence of a horizon.

For comparison, we also examine Killing trajectories, whose motion is generated by the stationary and axial symmetries of Kerr spacetime. For an equatorial trajectory with constant angular velocity $\omega$, the unnormalised tangent is proportional to $\partial_t+\omega\partial_\phi$. We define its squared norm by
\begin{equation}
    A^2 \equiv
    g\bigl(\partial_t+\omega\partial_\phi,\,
           \partial_t+\omega\partial_\phi\bigr),
    \label{eq:intro_A}
\end{equation}
where $\partial_t$ and $\partial_\phi$ are the coordinate vector fields associated with time translations and rotations. The corresponding unit tangent is obtained by dividing by $A$, and the trajectory is timelike when $A^2>0$. A surface on which $A^2=0$ is a light surface, where the Killing tangent becomes null. We derive closed expressions for $\kappa$ and $\tau_1$ and find that both diverge as $A^{-2}$ for the light-surface limits considered here. This behaviour provides a useful comparison with the null-endpoint limits of non-Killing trajectories (Sec.~\ref{sec:killing}).

Finally, we distinguish the first FS torsion from the rotation rate of a gyroscope relative to a regular geometrical frame. The first torsion describes rotation relative to the acceleration-adapted direction defined by the FS frame. This construction requires nonzero curvature; when $\kappa=0$, as for a geodesic, the acceleration no longer selects a preferred FS direction and the ordinary first torsion is undefined. We therefore also calculate the gyroscope's rotation rate relative to a frame that remains regular across the horizon. For the geodesic plunge considered here, this rate remains finite at horizon crossing and agrees with Marck's precession rate (Sec.~\ref{sec:obs}).

The results are restricted to the trajectory families and limiting conditions studied in this paper. They do not provide a classification of all possible worldlines or establish a general criterion for distinguishing black holes from horizonless compact objects.

The paper is organised as follows. Section~\ref{sec:setup} introduces the Kerr geometry and the FS invariants used throughout the analysis. Section~\ref{sec:killing} studies Killing trajectories, which provide a reference class associated with spacetime symmetries. Section~\ref{sec:charts} examines non-Killing spiral trajectories and tests the coordinate independence of their FS invariants. Section~\ref{sec:null} investigates the limiting behaviour of non-Killing trajectories approaching null endpoints. Section~\ref{sec:horizon} discusses horizon crossing and the extremal limit. Section~\ref{sec:obs} relates FS torsion to gyroscope rotation measured relative to a regular geometrical frame. Finally, Sec.~\ref{sec:disc} summarises the results and discusses the limitations of the analysis. Details of the derivations and numerical calculations are provided in the appendices.


\section{Geometry and Frenet--Serret formalism}
\label{sec:setup}

We use the metric signature $(+,-,-,-)$ and geometrized units, $G=c=1$. Unless stated otherwise, numerical results are given in units of $M=1$. In Boyer--Lindquist (BL) coordinates $(t,r,\theta,\phi)$, the Kerr metric is
\begin{align}
\Sigma &= r^2+a^2\cos^2\theta,
&
\Delta &= r^2-2Mr+a^2,
\label{eq:SigDel}\\
\dd s^2
&=
\left(1-\frac{2Mr}{\Sigma}\right)\dd t^2
+\frac{4Mar\sin^2\theta}{\Sigma}\dd t\,\dd\phi
-\frac{\Sigma}{\Delta}\dd r^2
-\Sigma\,\dd\theta^2
\nonumber\\
&\quad
-\sin^2\theta
\left(r^2+a^2+
\frac{2Ma^2r\sin^2\theta}{\Sigma}\right)\dd\phi^2.
\label{eq:Kerr}
\end{align}
Here, $M$ is the mass of the black hole and $a$ is its spin parameter. The functions $\Sigma$ and $\Delta$ are standard abbreviations used to express the metric compactly. For a sub-extremal Kerr black hole, $|a|<M$, the outer and inner horizons are located at
\begin{equation}
r_\pm=M\pm\sqrt{M^2-a^2}.
\label{eq:horizons}
\end{equation}
The outer horizon is at $r=r_+$, while $r=r_-$ denotes the inner horizon.

\subsection{Frenet--Serret scalars}
\label{subsec:FS}

We first define the Frenet--Serret (FS) quantities used to describe a timelike worldline. Let $u^\mu$ be its unit tangent, normalised according to
$u^\mu u_\mu=1$. The four-acceleration is
\begin{equation}
A^\mu \equiv u^\nu\nabla_\nu u^\mu,
\label{eq:proper_acceleration}
\end{equation}
where $\nabla_\nu$ denotes the covariant derivative associated with the spacetime metric. The FS curvature $\kappa$ is the magnitude of the four-acceleration:
\begin{equation}
\kappa \equiv \sqrt{-A_\mu A^\mu}.
\label{eq:kappa_def}
\end{equation}
Thus, $\kappa$ measures the proper acceleration required to maintain the given trajectory.

When $\kappa\neq0$, the first spatial vector of the FS frame is determined by the direction of the four-acceleration,
\begin{equation}
e_1^\mu=\frac{A^\mu}{\kappa}.
\end{equation}
The first torsion $\tau_1$ describes the rotation of the acceleration-adapted FS frame along the worldline. More precisely, it measures the part of the covariant derivative of $e_1^\mu$ that is orthogonal to the timelike tangent and the acceleration direction. It can be defined through
\begin{equation}
\tau_1^2
=
-\left(\frac{D e_1^\mu}{\dd\tau}-\kappa u^\mu\right)
 \left(\frac{D e_{1\mu}}{\dd\tau}-\kappa u_\mu\right),
\label{eq:tau1def}
\end{equation}
where $D/\dd\tau\equiv u^\nu\nabla_\nu$ denotes covariant differentiation along the worldline.

The FS frame vectors obey the transport equations~\cite{IV}
\begin{align}
\frac{D e_0^\mu}{\dd\tau}
&=\kappa e_1^\mu,
&
\frac{D e_1^\mu}{\dd\tau}
&=\kappa e_0^\mu+\tau_1e_2^\mu,
\nonumber\\
\frac{D e_2^\mu}{\dd\tau}
&=-\tau_1e_1^\mu+\tau_2e_3^\mu,
&
\frac{D e_3^\mu}{\dd\tau}
&=-\tau_2e_2^\mu,
\label{eq:FS}
\end{align}
where $e_0^\mu=u^\mu$, and $\tau_2$ is the second torsion. For a trajectory confined to the equatorial plane of a reflection-symmetric spacetime, $\tau_2=0$.

The FS construction requires special care when the curvature vanishes. If $\kappa=0$, the acceleration no longer determines a preferred direction for $e_1^\mu$, and the standard FS frame becomes degenerate. In particular, along an exact geodesic, the ordinary FS torsion is not defined by this construction. Gyroscope transport along such a trajectory must instead be described using parallel transport or another suitably defined reference frame.

For practical calculations, $\tau_1$ can also be expressed in terms of the four-acceleration and its covariant derivative. Define
\begin{equation}
B^\mu\equiv\frac{D A^\mu}{\dd\tau},
\qquad
\dot{\kappa}\equiv\frac{\dd\kappa}{\dd\tau}.
\end{equation}
Using $\kappa^2=-A_\mu A^\mu$ and the FS transport equations, one obtains
\begin{equation}
\tau_1^2
=
\frac{\kappa^4-\dot{\kappa}^{\,2}-B_\mu B^\mu}
{\kappa^2}.
\label{eq:tau1B}
\end{equation}
All derivatives in these expressions are covariant derivatives along the worldline. Consequently, differentiating vector components alone is insufficient: the connection terms involving the Christoffel symbols must also be included. When the tangent depends explicitly on the radial coordinate, its radial variation along the trajectory must likewise be accounted for through the chain rule.

\subsection{Non-unit tangents and an exact curvature identity}
\label{subsec:nonunit}

For prescribed trajectories, it is often convenient to work with a tangent vector that is not normalised to unit proper time. Let
\begin{equation}
v^\mu\equiv\frac{\dd x^\mu}{\dd s},
\end{equation}
where $s$ is a parameter along the curve. We introduce the squared norm $S$, the covariant acceleration $q^\mu$ associated with this parameter, and the scalar $Q$:
\begin{equation}
S\equiv g_{\mu\nu}v^\mu v^\nu,
\qquad
q^\mu\equiv v^\nu\nabla_\nu v^\mu,
\qquad
Q\equiv q_\mu v^\mu
=\frac{1}{2}\frac{\dd S}{\dd s}.
\label{eq:SqQ}
\end{equation}
With our metric signature, the curve is timelike wherever $S>0$ and approaches a null direction when $S\to0^+$. On a timelike segment, the unit tangent and proper-time element are related to the non-unit tangent by
\begin{equation}
u^\mu=\frac{v^\mu}{\sqrt{S}},
\qquad
\dd\tau=\sqrt{S}\,\dd s.
\label{eq:proper_time_nonunit}
\end{equation}

These relations allow the proper four-acceleration to be expressed directly in terms of $v^\mu$, $q^\mu$, $S$, and $Q$. Substituting the resulting expression into $\kappa^2=-A_\mu A^\mu$ gives the exact identity
\begin{equation}
A^\mu=\frac{q^\mu}{S}-\frac{Q}{S^2}v^\mu,
\qquad
\kappa^2=\frac{Q^2}{S^3}-\frac{q_\mu q^\mu}{S^2}.
\label{eq:kappaS}
\end{equation}
Here, $A^\mu$ again denotes the proper four-acceleration defined in Eq.~\eqref{eq:proper_acceleration}; it is not the normalisation factor used later for Killing trajectories.

Equation~\eqref{eq:kappaS} is valid for any smooth parametrised timelike curve with $S>0$. It is particularly useful near a null endpoint, where $S\to0^+$. The identity shows that the limiting behaviour of $\kappa$ depends on how $S$, $Q$, and $q_\mu q^\mu$ behave. In particular, if $Q\to Q_0\neq0$ while $q_\mu q^\mu$ remains finite, the first term dominates and the curvature satisfies
\begin{equation}
\kappa\sim |Q_0|S^{-3/2}.
\label{eq:kappa_null_scaling}
\end{equation}
This scaling is conditional rather than universal: other behaviours are possible if the stated assumptions do not hold. We use this identity in Sec.~\ref{sec:null} to analyse the curvature near null endpoints and to derive the conditions under which the first torsion can approach a finite limit.

\section{Killing observers}
\label{sec:killing}

We begin with trajectories generated by the symmetries of Kerr spacetime. These Killing trajectories provide a useful reference for the more general non-Killing motion considered later. For equatorial stationary observers, the angular velocity is constant along each orbit, and the unit tangent can be written as $u=(1,0,0,\omega)/A$, where $A^2=g(\partial_t+\omega\partial_\phi,\partial_t+\omega\partial_\phi)$. The condition $A^2>0$ ensures that the orbit is timelike. Because the norm of the corresponding unnormalised Killing tangent is constant along each orbit, the Frenet--Serret curvature and first torsion have particularly simple expressions. We first derive these quantities and examine their behaviour at light surfaces, where the stationary trajectory becomes null. We then discuss allowed angular velocities and the limiting behaviour of Killing observers near sub-extremal and extremal horizons.

The normalization factor is determined by
\begin{equation}
A^2=1-(r^2+a^2)\omega^2
-\frac{2M(1-a\omega)^2}{r}.
\end{equation}
The condition $A^2>0$ ensures that the orbit is timelike. Since the norm of the unnormalised Killing tangent is constant along each orbit, the curvature identity simplifies to $\kappa^2=-q\cdot q/A^4$. Setting $M=1$, we obtain the following closed expressions for the FS curvature and first torsion:
\begin{align}
\kappa&=\frac{\sqrt{\Delta}\,
\bigl|r^3\omega^2-(1-a\omega)^2\bigr|}
{r^3A^2},\label{eq:kapK}\\
\tau_1&=\frac{\bigl|a(1+\omega^2(a^2+3r^2))
-2a^2\omega+\omega r^2(r-3)\bigr|}
{r^3A^2}.\label{eq:tauK}
\end{align}
The curvature expression in Eq.~\eqref{eq:kapK} reproduces the result for $\kappa^2$ given in Eq.~(12) of the first version of this work \cite{prev}. We have also checked the torsion expression in Eq.~\eqref{eq:tauK} against a direct evaluation of the FS equations; the details are given in Appendix~\ref{app:num}.

The two expressions have a further useful relation. Denoting their numerators by $N_\kappa$ and $N_\tau$, respectively, we find
\begin{equation}
\kappa^2-\tau_1^2
=-\frac{P(r,a,\omega)}{r^4A^2},
\label{eq:diffK}
\end{equation}
where
\begin{equation}
P=\omega^2r^4-2\omega^2r^3-2a^2\omega^2r
-a^2\omega^2+2a\omega r+2a\omega-1.
\end{equation}
This relation has been verified symbolically and is useful for understanding the behaviour of the two invariants near a light surface.

\paragraph*{Light surfaces.}
A light surface is reached when $A^2$ vanishes, so that the stationary orbit becomes null. Consider a light surface at which $\Delta>0$ and the numerators in Eqs.~\eqref{eq:kapK} and \eqref{eq:tauK} remain nonzero. Both $\kappa$ and $\tau_1$ then diverge as $A^{-2}$. Equation~\eqref{eq:diffK} shows, however, that their squared difference grows only as $A^{-2}$. Consequently, their ratio approaches unity, while the difference $\kappa-\tau_1$ remains finite:
\begin{equation}
\frac{\kappa}{\tau_1}\longrightarrow 1,
\qquad
\kappa-\tau_1\longrightarrow
-\frac{P}{2r|N_\tau|}.
\end{equation}
The light-surface examples in Table~\ref{tab:light} illustrate this behaviour. In each case, the magnitudes of the two numerators agree at the light surface, $|N_\kappa|=|N_\tau|$.

These light surfaces can occur outside the event horizon, and they are also present in the super-extremal cases considered here. This distinction is important when interpreting divergences found near the horizon in BL coordinates. For example, when $a=0.5$ and $\omega=0.15$, the light surface lies at $r_\ell=1.86855$, just outside the horizon at $r_+=1.86603$. The divergence along this family of Killing orbits is associated with the orbit becoming null at $r_\ell$, rather than with the horizon itself.

\begin{table}[b]
\caption{Light surfaces of equatorial Killing observers ($M=1$). The table gives the light-surface radius $r_\ell$ and the ratio $|N_\kappa|/|N_\tau|$ for the numerators in Eqs.~\eqref{eq:kapK} and \eqref{eq:tauK}.}
\label{tab:light}
\begin{ruledtabular}
\begin{tabular}{cccc}
$a$ & $\omega$ & $r_\ell$ & $|N_\kappa|/|N_\tau|$\\
\hline
0.5 & 0.15 & 1.8685509 & 1.0\\
1.0 & 0.15 & 1.5667929 & 1.0\\
1.5 & 0.25 & 3.1218533 & 1.0\\
0.5 & $-0.15$ & 2.8458304 & 1.0\\
0.9 & 0.2 & 1.541152 & 1.0\\
\end{tabular}
\end{ruledtabular}
\end{table}

\begin{figure*}[t]
\centering
\begin{minipage}[t]{0.48\textwidth}
  \centering
  \includegraphics[width=\linewidth]{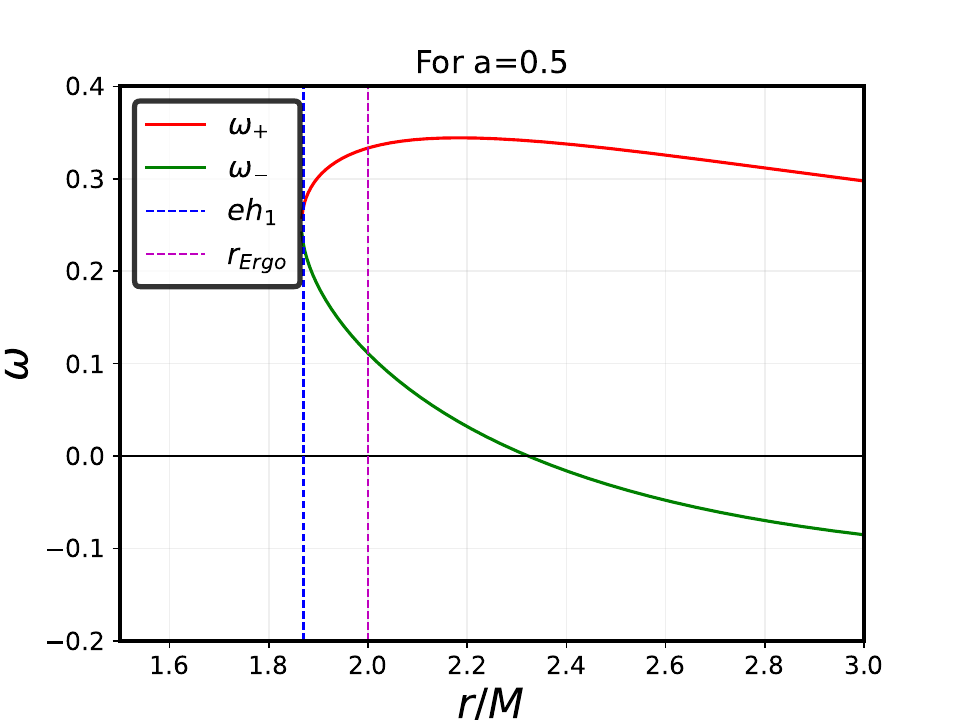}
  \vspace{1mm}\small (a) $a/M=0.5$
\end{minipage}\hfill
\begin{minipage}[t]{0.48\textwidth}
  \centering
  \includegraphics[width=\linewidth]{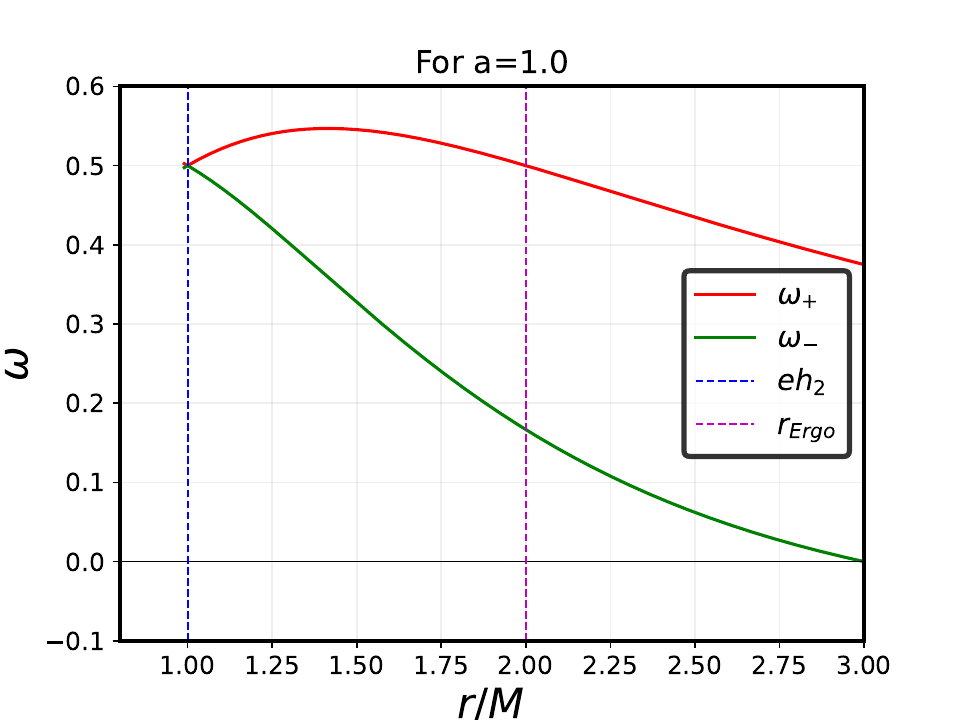}
  \vspace{1mm}\small (b) $a/M=1$
\end{minipage}
\caption{Allowed angular-velocity roots $\omega_\pm$ for equatorial stationary observers. The horizon and ergosurface are marked in the source plots. The ZAMO angular velocity lies between the two roots and, when a horizon exists, approaches the horizon angular velocity there.}
\label{fig:zamo-roots}
\end{figure*}

The allowed angular velocities of stationary observers also help clarify which circular trajectories are timelike. At a given radius, the condition $A^2>0$ restricts $\omega$ to lie between two limiting values, $\omega_-$ and $\omega_+$. Figure~\ref{fig:zamo-roots} shows these roots for two values of the spin parameter. The zero-angular-momentum observer (ZAMO) angular velocity lies between the roots and approaches the horizon angular velocity when a horizon is present.

\paragraph*{Horizon-corotating observers.}
We next consider the angular velocity
$\omega_H=a/(r_+^2+a^2)$ associated with the horizon. For a sub-extremal black hole, we find numerically that, as $r\to r_+$,
\begin{equation}
\kappa A\longrightarrow
\kappa_{\rm s}=\frac{r_+-r_-}{2(r_+^2+a^2)}.
\end{equation}
For example, at $a=0.5$, the ratio $\kappa A/\kappa_{\rm s}$ is $0.9999975$ at $r-r_+=10^{-6}$. This is consistent with the near-horizon acceleration behaviour familiar from Rindler-type analyses.

The torsion behaves differently in this limit. The numerator $N_\tau$ vanishes at $(r_+,\omega_H)$, as checked numerically for $a=0.3$, $0.5$, and $0.9$, and $\tau_1$ tends to zero linearly with $r-r_+$. We find the same behaviour in the Kerr--Newman example with $Q^2=0.25$ and $a=0.5$: the ratio $\kappa A/\kappa_{\rm s}$ reaches $0.9999999$ at $r-r_+=10^{-7}$, while $\tau_1$ tends to zero. These numerical results suggest that the behaviour may extend to a broader class of Killing horizons, although a general proof is beyond the scope of this work.

\paragraph*{Extremal Kerr.}
The extremal case requires separate treatment. For $a=M=1$ and $\omega=1/2$, the norm of the horizon-corotating tangent is
\begin{equation}
A^2=-\frac{(r-1)^2(r+2)}{4r}<0
\end{equation}
for every $r\neq1$. Thus, this particular tangent is spacelike away from the horizon and cannot be used as a timelike family to take the near-horizon limit discussed above.

For angular velocities below the extremal horizon value, $\omega<1/2$, light surfaces instead approach $r=1$ as $\omega$ tends to $1/2$. Representative values are
\begin{equation}
\begin{aligned}
(\omega,r_\ell,\kappa A^2=\tau_1 A^2)
&=(0.45,\,1.2222,\,8.18\times10^{-3}),\\
&\quad (0.49,\,1.0134,\,1.32\times10^{-4}),\\
&\quad (0.499,\,1.0013,\,1.33\times10^{-6}).
\end{aligned}
\end{equation}
These results illustrate how the light surface approaches the extremal horizon in this limit. They should not, however, be interpreted as establishing the same horizon-corotating limit for a timelike family at exact extremality.

\section{Non-Killing worldlines and coordinate charts}
\label{sec:charts}

The Killing trajectories considered in the preceding section form a special class of worldlines associated with spacetime symmetries. A general accelerated trajectory need not possess this symmetry and may have a nonzero radial velocity. We now consider prescribed non-Killing trajectories in Kerr spacetime, using spiral curves as illustrative examples. The purpose is twofold: to examine the Frenet--Serret invariants along these trajectories and to establish how a worldline must be represented consistently when changing coordinates.

\begin{figure}[t]
\centering
\includegraphics[width=0.9\columnwidth]{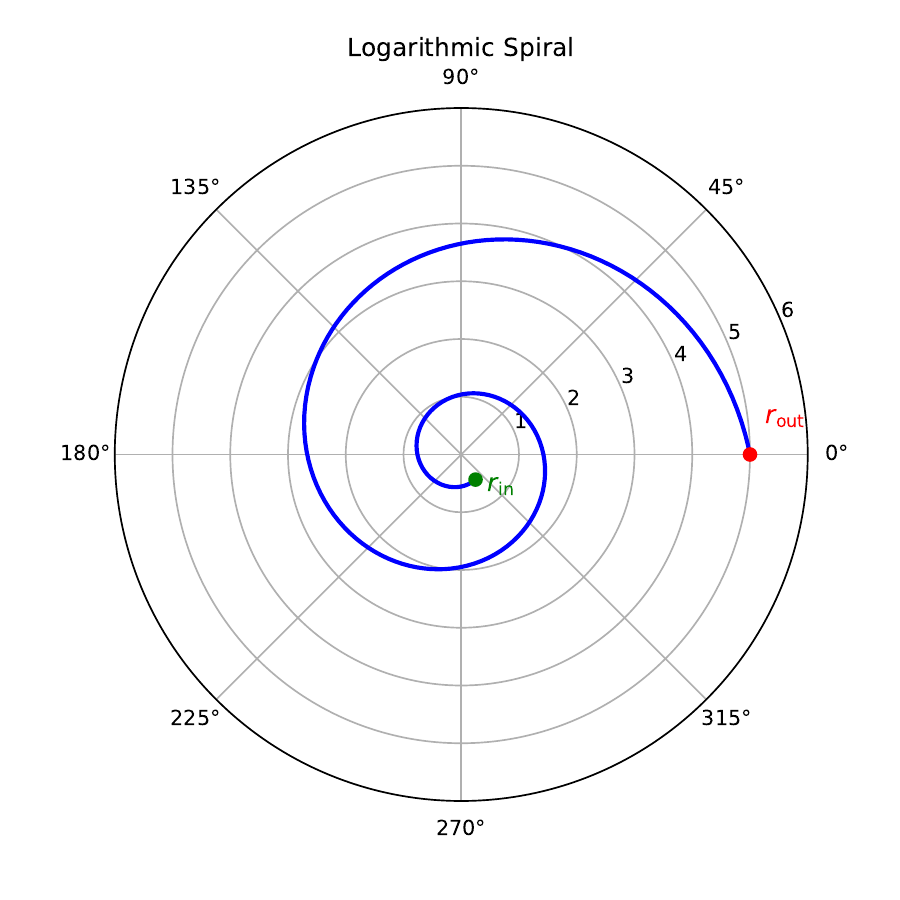}
\caption{Schematic logarithmic spiral used as a kinematic model for a non-Killing trajectory. The figure illustrates the path geometry only; it is not a self-force-driven inspiral.}
\label{fig:spiral-path}
\end{figure}

Figure~\ref{fig:spiral-path} shows a schematic logarithmic spiral to convey the geometry of this class of motion. The spiral trajectory is defined by the equation $r=r_{out} e^{\lambda\phi}$, where $\lambda$ is the spiral path parameter. It is intended only as a kinematic illustration; it is not the trajectory used in the calculations below and does not represent a self-force-driven inspiral. Our quantitative analysis instead uses explicitly prescribed tangent fields, whose Frenet--Serret invariants can be evaluated directly.

\subsection{Spiral trajectories and their coordinate representations}

To compare the same physical trajectory in different coordinate systems, the tangent to the worldline must be transformed along with the coordinates. We begin with an ingoing Kerr--Schild-type chart $(t',r,\theta,\phi')$, related to Boyer--Lindquist (BL) coordinates by
\begin{equation}
\dd t'=\dd t+\frac{2Mr}{\Delta}\dd r,\qquad
\dd\phi'=\dd\phi+\frac{a}{\Delta}\dd r.
\label{eq:KS}
\end{equation}
The coordinates $r$ and $\theta$ are unchanged. We have checked numerically that the metric obtained by transforming the BL metric agrees with its Kerr--Schild form. Under this coordinate transformation, the components of a tangent transform according to
\begin{equation}
u'^t=u^t+\frac{2Mr}{\Delta}u^r,\quad
u'^r=u^r,\quad
u'^\phi=u^\phi+\frac{a}{\Delta}u^r.
\label{eq:uKS}
\end{equation}

As a concrete example, consider an equatorial spiral specified in the ingoing chart by the unnormalised tangent
\begin{equation}
v'^\mu=(1,\ \lambda\omega r,\ 0,\ \omega),
\label{eq:vKS}
\end{equation}
where $\lambda$ and $\omega$ are constants. Transforming this tangent to BL coordinates gives
\begin{equation}
v^\mu_{\rm BL}=
\Bigl(1-\frac{2Mr^2\lambda\omega}{\Delta},\
\lambda\omega r,\ 0,\
\omega-\frac{a\lambda\omega r}{\Delta}\Bigr).
\label{eq:vBL}
\end{equation}
The time and azimuthal components in BL coordinates diverge as $\Delta\to0$. This behaviour, by itself, does not imply a divergence in the FS curvature or torsion: these are scalars associated with the physical worldline, not with the coordinate components used to describe it.

The distinction becomes important if the same component prescription is imposed independently in the two coordinate systems. In particular, taking
$v^\mu=(1,\lambda\omega r,0,\omega)$ directly in BL coordinates defines a different curve from the one specified by Eq.~\eqref{eq:vKS}. Transforming the latter curve according to Eq.~\eqref{eq:uKS} does not, in general, reproduce the former prescription. Consequently, any difference between the FS invariants of these two curves reflects a difference in their trajectories, rather than a failure of coordinate invariance. For one and the same smooth worldline, the metric, tangent, and connection must all be transformed consistently, and the resulting values of $S$, $\kappa$, and $\tau_1$ must agree wherever the quantities are defined.

\subsection{The BL-form curve cannot reach the horizon}

The difference between the two prescriptions can also be understood by examining the norm of the tangent. For the curve defined directly in BL coordinates, the squared norm $S_{\rm BL}$ contains the term
$-\Sigma(\lambda\omega r)^2/\Delta$. As $\Delta\to0^+$, this term tends to $-\infty$, while the remaining terms stay finite. Thus, for $\lambda\omega\neq0$, the curve becomes null at some radius $r_\ell>r_+$ and is spacelike closer to the horizon. It cannot describe a timelike observer all the way to the horizon.

\begin{figure}[t]
\includegraphics[width=\columnwidth]{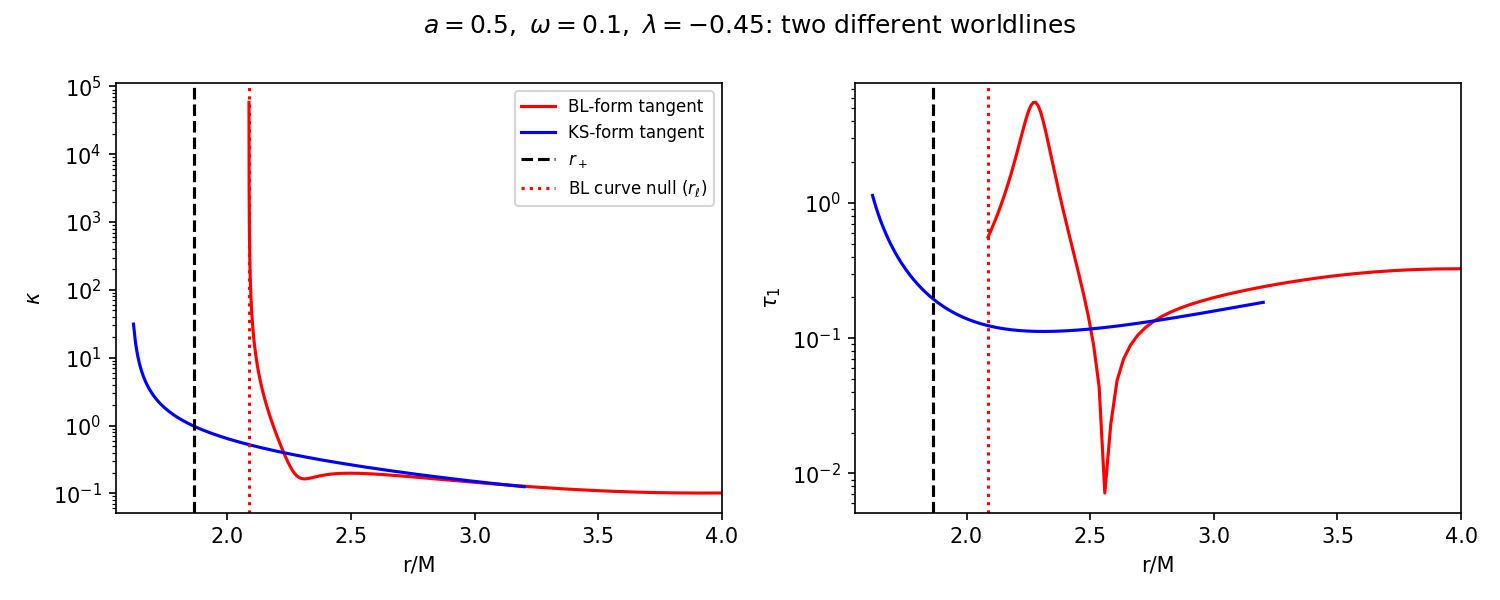}
\caption{The FS curvature $\kappa$ and first torsion $\tau_1$ for two different worldlines with the same parameters $(a,\omega,\lambda)=(0.5,0.1,-0.45)$. The red curves correspond to the tangent $(1,\lambda\omega r,0,\omega)$ prescribed directly in BL coordinates, while the blue curves correspond to the tangent prescribed in the ingoing Kerr--Schild chart. The BL-form curve becomes null at $r_\ell=2.087>r_+$ (dotted line), where $\kappa$ diverges. The Kerr--Schild curve crosses the horizon at $r_+=1.866$ (dashed line) with finite invariants and becomes null at approximately $r=1.606$.}
\label{fig:two}
\end{figure}

For $a=0.5$, $\omega=0.1$, and $\lambda=-0.45$, the null radius is $r_\ell=2.0869$, outside the horizon at $r_+=1.8660$. The divergence of the FS curvature as this curve approaches $r_\ell$ should therefore not be interpreted as a horizon divergence. Instead, it occurs as the prescribed BL-form tangent approaches a null direction. Figure~\ref{fig:two} illustrates the difference between this curve and the spiral defined in the ingoing Kerr--Schild chart. The latter crosses the horizon with finite FS invariants and reaches its own null endpoint at approximately $r=1.606$.

\subsection{Numerical test of chart independence}

We next test explicitly whether the FS invariants agree when a single worldline is represented in different coordinate systems. We calculate $S$, $\kappa^2$, and $\tau_1^2$ using Eqs.~\eqref{eq:tau1def}--\eqref{eq:tau1B}, evaluating the Christoffel symbols and carrying out high-precision numerical differentiation along the curve. The numerical procedure is described in Appendix~\ref{app:num}.

In addition to BL coordinates, we consider three horizon-regular charts related to BL by transformations of the form
$\dd t'=\dd t+F\,\dd r$ and
$\dd\phi'=\dd\phi+G\,\dd r$ on the equatorial plane. These are the Kerr--Schild (KS) chart, with $F=2Mr/\Delta$ and $G=a/\Delta$; the ingoing Kerr (IK) chart, with $F=(r^2+a^2)/\Delta$ and $G=a/\Delta$; and a Doran-type (DO) chart, with
$F=\sqrt{2Mr(r^2+a^2)}/\Delta$ and
$G=(a/\Delta)\sqrt{2Mr/(r^2+a^2)}$. The metric components in these charts remain finite at the outer horizon.

We also test a transformation that depends explicitly on the polar angle,
\begin{equation}
t''=t'+\epsilon_1r\cos^2\theta,\qquad
\phi''=\phi'+\epsilon_2r\cos^2\theta,
\end{equation}
where $\epsilon_1=3/10$ and $\epsilon_2=1/5$. This provides a test beyond coordinate transformations that depend only on $r$. For this case, we use an off-equatorial trajectory whose tangent in the KS chart is
$(1,\lambda\omega r,\mu,\omega)$, with $\mu=0.03$, evaluated at $\theta=1$ rad.

In every comparison, the worldline is specified in one chart and then transformed to the other charts using the appropriate Jacobian. We do not independently impose the same tangent components in each coordinate system. As summarised in Table~\ref{tab:chart}, the calculated values of the FS invariants agree to all displayed digits in the charts that are regular at the evaluation point. This agreement holds for both equatorial and off-equatorial examples, including points inside the horizon where the corresponding coordinates remain regular. The final row of the table shows the BL-form tangent discussed above. Its different invariant values are expected because it represents a different worldline.

\begin{table}[t]
\caption{FS invariants of a single worldline evaluated in different coordinate charts, with $M=1$. For each row, the same trajectory is transformed consistently between the listed charts. The final row uses the BL-form tangent of Sec.~\ref{sec:charts}, which defines a different worldline.}
\label{tab:chart}
\begin{ruledtabular}
\begin{tabular}{lccccllll}
Curve & $a$ & $\omega$ & $\lambda$ & $r$ & $S$ & $\kappa^2$ & $\tau_1^2$ & Charts\\
\hline
equatorial & 0.5 & 0.1 & $-0.45$ & 2.5 & 0.34996875 & 0.0692327352084786 & 0.0136248281803597 & BL, KS, IK, DO\\
equatorial & 0.5 & 0.1 & $-0.45$ & 1.8\footnote{Inside the horizon, $r_+=1.866$.} & 0.111371222222222 & 1.70263947002147 & 0.0639969290288769 & KS, IK, DO\\
equatorial & 1.0 & 0.3 & $-0.45$ & 1.2 & 0.174549333333333 & 1.02519255323825 & 1.62662301579717 & BL, KS, IK, DO\\
equatorial\footnote{$a>M$: no horizon.} & 1.5 & 0.3 & $-0.15$ & 1.2 & 0.206357333333333 & 0.111770249667467 & 1.47497631938096 & BL, KS, IK, DO\\
off-equatorial & 0.5 & 0.1 & $-0.45$ & 2.5 & 0.3535123326032 & 0.071842659988373 & 0.0045056475827432 & BL, KS, KT\\
off-equatorial & 0.5 & 0.1 & $-0.45$ & 1.8 & 0.11162848924396 & 1.6400699853524 & 0.047238718038659 & KS, KT\\
\hline
BL-form (different curve) & 0.5 & 0.1 & $-0.45$ & 2.5 & 0.16026562 & 0.039250267 & 0.015364477 & BL\\
\end{tabular}
\end{ruledtabular}
\end{table}

\subsection{Scope: other horizon-regular charts}

The coordinate-invariance argument applies to any valid coordinate transformation, provided the metric and the worldline are transformed consistently. It is nevertheless useful to test this principle in explicit coordinate systems, particularly when their transformations have a more complicated dependence on spacetime position.

One possible extension would be to repeat the calculation in the double-null Kerr coordinates constructed by Arga\~naraz and Moreschi \cite{AM}. Their transformation is built from null surfaces and depends on both $r$ and $\theta$. We have not implemented that transformation, and therefore make no claim to have tested the invariants numerically in those coordinates. The present results are restricted to the charts listed above, including a $\theta$-dependent transformation, and to the equatorial and off-equatorial examples described in this section.


The preceding analysis establishes how non-Killing trajectories must be specified consistently across coordinate charts and how their Frenet--Serret invariants can be compared. We now investigate a different aspect of these trajectories: the limiting behaviour of the invariants when a timelike curve approaches a null endpoint. The endpoint need not coincide with the event horizon, so its properties must be analysed independently of whether a horizon is present.

\section{Null endpoints of non-Killing curves}
\label{sec:null}

\subsection{Curvature}

Consider a smooth timelike curve that approaches a null endpoint at $s=s_0$, so that $S\to0^+$. If $Q$ approaches a nonzero value $Q_0$ and $q\!\cdot\!q$ stays finite, Eq.~\eqref{eq:kappaS} gives
\begin{equation}
\kappa^2=Q_0^2S^{-3}[1+o(1)],\qquad \kappa\sim|Q_0|\,S^{-3/2}.\label{eq:kapasym}
\end{equation}
The condition $Q_0\neq0$ is essential. For a Killing orbit $S$ is constant along the curve ($Q=0$), the endpoint is approached through a family of orbits, and $\kappa\sim S^{-1}$ instead (Sec.~\ref{sec:killing}).

\subsection{Torsion}

\begin{proposition}\label{prop:tau}
Let $x(s)$ be a smooth curve with tangent $v=\dd x/\dd s$, $S=g(v,v)$, $q=D_vv$, $p=D_vq$, and suppose $S>0$ on one side of $s_0$, $S(s_0)=0$ and
$n_1\equiv S'(s_0)=2Q_0\neq0$. Then $\kappa=|S'|/(2S^{3/2})\,[1+O(S)]$ and $\tau_1$ has the finite limit
\begin{equation}
\ts{}^2=\frac{4\bigl[-4B_0^3+4B_0^2n_2-2B_0B_1n_1-B_0n_2^2+B_1n_1n_2-D_0n_1^2\bigr]}{n_1^4},\label{eq:tauStar}
\end{equation}
where $n_2=S''(s_0)$, $B_0=q\!\cdot\! q$, $B_1=2\,q\!\cdot\! p$ and $D_0=p\!\cdot\! p$, all evaluated at $s_0$.
\end{proposition}

The proof is a series expansion of Eq.~\eqref{eq:tau1B} written entirely in terms of $S$, $q\!\cdot\! q$ and $p\!\cdot\! p$ (Appendix~\ref{app:der}); the leading divergences of order $S^{-6}$, $S^{-5}$ and $S^{-4}$
cancel identically, leaving a finite numerator at order $S^{-3}\sim\kappa^2$. The bracket is non-negative for any curve to which the hypotheses apply.
The proposition only requires data at the endpoint; it shows that curvature and torsion behave differently at a null endpoint, and that the torsion limit depends on the
local geometry of the curve (through $q$, $p$ and $S''$) and not on the presence of a horizon.

\subsection{Validation}

\begin{figure}[t]
\includegraphics[width=\columnwidth]{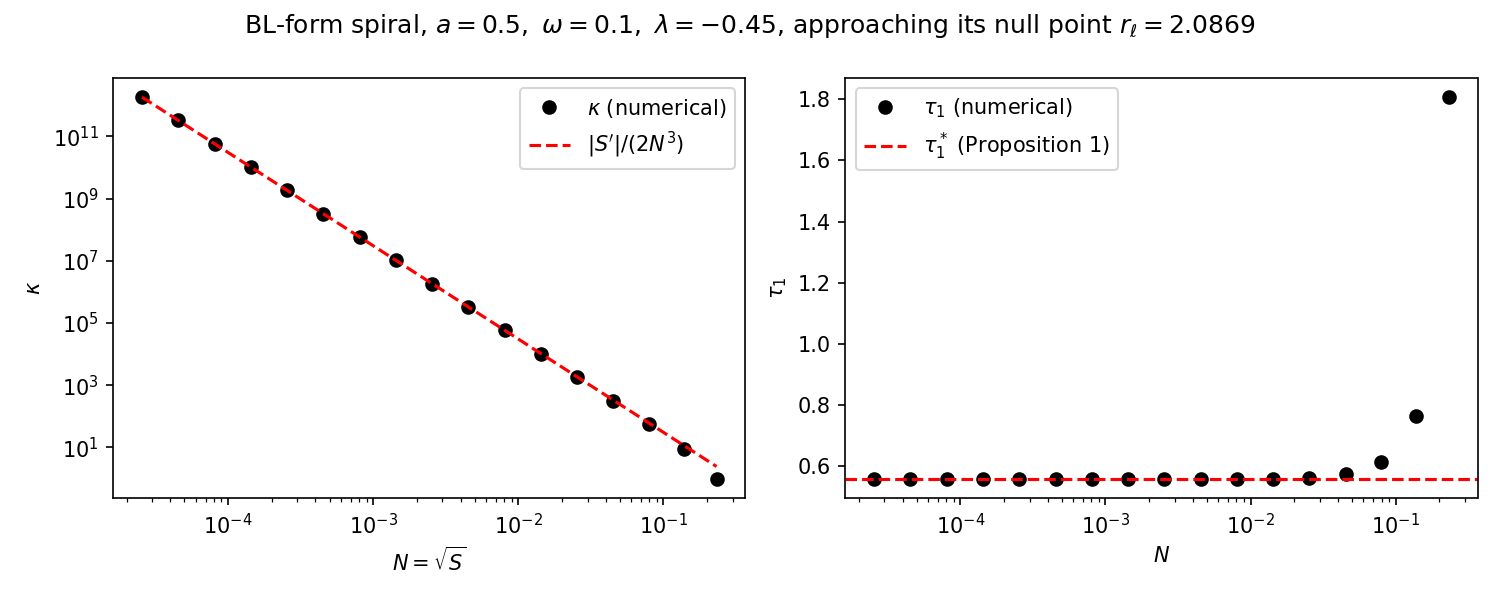}
\caption{Approach to the null point for the BL-form spiral $(a,\omega,\lambda)=(0.5,0.1,-0.45)$. Left: $\kappa$ against $N=\sqrt{S}$ with the asymptote $|S'|/(2N^3)$.
Right: $\tau_1$ against $N$ with the analytic limit $\tau_1^*$ of Eq.~\eqref{eq:tauStar}.}
\label{fig:null}
\end{figure}

Table~\ref{tab:null} compares Eq.~\eqref{eq:tauStar} with numerical limits of $\tau_1$. The last two rows are the prescribed Kerr--Schild spirals of Sec.~\ref{sec:examples}, with
$(a/M,\omega M,\lambda)=(1/2,\,1/10,\,-9/20)$ and $(3/2,\,3/10,\,-9/20)$; the others are BL-form spirals approached from the timelike side. The curvature law Eq.~\eqref{eq:kapasym} holds to at least four digits in all nine
cases scanned (six to seven digits in most), and $\tau_1$ approaches the finite value of Eq.~\eqref{eq:tauStar} with relative differences below $10^{-5}$ (the residuals come from the finite distance to the endpoint, $10^{-8}$ or $10^{-9}$).
An off-equatorial BL-form curve ($a=0.5$, $\theta=1$ rad, $\mu=0.03$) gives $\kappa S^{3/2}=0.0300578$ against $|S'|/2=0.0300579$, with $\tau_1\to0.4983$ finite.
Figure~\ref{fig:null} shows both asymptotics for the first row.

\begin{table}[t]
\caption{Null-endpoint limits: $|Q_0|=|S'|/2$, the analytic torsion limit $\tau_1^{*2}$ from Eq.~\eqref{eq:tauStar}, and the numerical $\tau_1^2$ at distance $10^{-8}$ (BL-form) or $10^{-9}$ (KS-form) from the endpoint ($M=1$).}
\label{tab:null}
\begin{ruledtabular}
\begin{tabular}{llccccc}
Tangent form & $(a,\omega,\lambda)$ & $r_{\rm null}$ & $|Q_0|$ & $\tau_1^{*2}$ (analytic) & $\tau_1^2$ (numerical)\\
\hline
BL-form & $(0.5,\,0.1,\,-0.45)$ & 2.0869213 & 0.0305546 & 0.31157967 & 0.31157973\\
BL-form & $(0.5,\,0.1,\,-0.2)$ & 1.9550729 & 0.0153255 & 1.5043998 & 1.5044005\\
BL-form & $(0.9,\,0.1,\,-0.3)$ & 1.7644520 & 0.0140881 & 0.76773028 & 0.76772990\\
BL-form & $(1.0,\,0.09,\,-0.2)$ & 1.7208239 & 0.0082684 & 261.29623 & 261.29578\\
BL-form & $(1.5,\,0.1,\,-0.3)$ & 1.5190186 & 0.0134367 & 6346.5410 & 6346.5161\\
BL-form & $(0.3,\,0.2,\,-0.3)$ & 1.7825203 & 0.0710986 & 3.3627550 & 3.3627540\\
BL-form & $(0,\,0.1,\,-0.3)$ & 2.2101488 & 0.0173217 & 3.3339532 & 3.3339537\\
KS-form & $(0.5,\,0.1,\,-0.45)$ & 1.6062405 & 0.0235791 & 2.0367632 & 2.0367631\\
KS-form & $(1.5,\,0.3,\,-0.45)$ & 2.0945981 & 0.0669359 & 11.478773 & 11.478772\\
\end{tabular}
\end{ruledtabular}
\end{table}

\subsection{Exact symbolic examples}
\label{sec:examples}

\begin{figure}[t]
\includegraphics[width=\columnwidth]{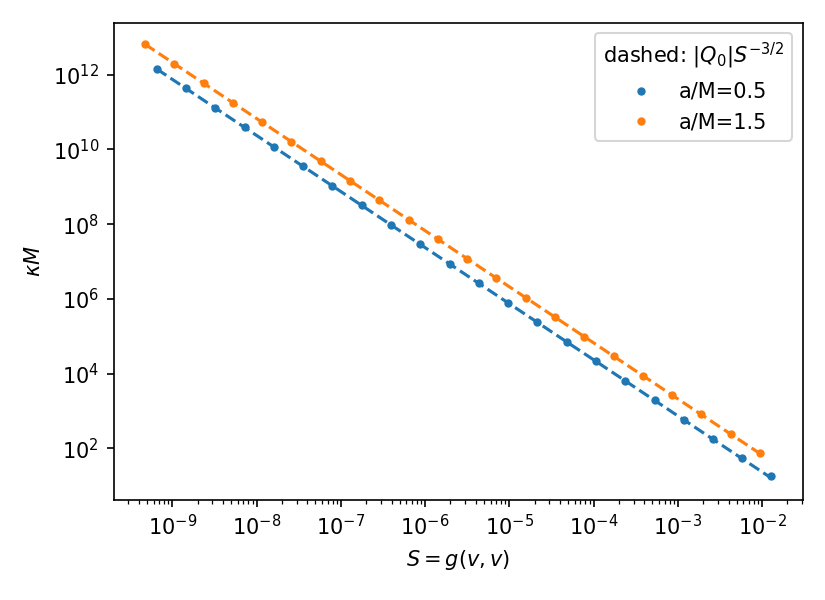}
\caption{FS curvature against $S$ for the two Kerr--Schild spirals; dashed lines are $|Q_0|S^{-3/2}$ (Eq.~\eqref{eq:kapasym}).}
\label{fig:curv}
\end{figure}

\begin{figure}[t]
\includegraphics[width=\columnwidth]{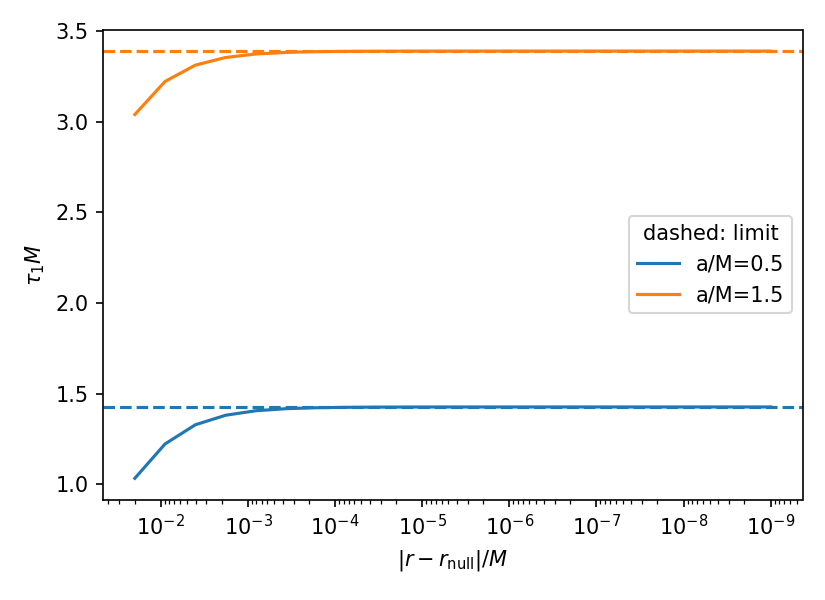}
\caption{$\tau_1$ against distance from the null endpoint for the same curves; dashed lines are the limits $1.4271521$ and $3.3880337$.}
\label{fig:tors}
\end{figure}

\begin{figure}[t]
\includegraphics[width=\columnwidth]{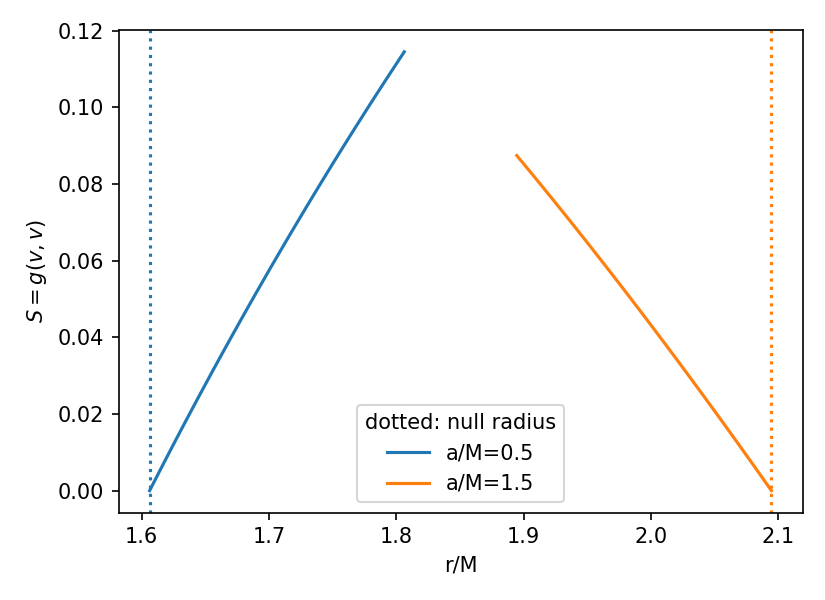}
\caption{$S=g(v,v)$ on the timelike side of each null endpoint. The $a/M=0.5$ root lies inside the outer horizon; $a/M=1.5$ has no horizon.}
\label{fig:dom}
\end{figure}

For the two Kerr--Schild spirals above, the exact equatorial Kerr--Schild metric and exact rational parameters give $\tau_1^2(r)$ as a rational function. For $a/M=3/2$,
$\tau_1^2=9P_{3/2}(r)^2/[4r^4R_{3/2}(r)^2]$, and for $a/M=1/2$
\begin{equation}
\tau_1^2=\frac{P_{1/2}(r)^2}{4r^4R_{1/2}(r)^2},\label{eq:rational}
\end{equation}
with the integer polynomials of Appendix~\ref{app:poly}. The denominators do not vanish at the null roots, which gives $\tau_1\to1.4271521$ ($a/M=0.5$, from $r\to r_{\rm null}^+$) and
$\tau_1\to3.3880337$ ($a/M=1.5$, from $r\to r_{\rm null}^-$), with curvature coefficients $\kappa S^{3/2}\to0.0235791$ and $0.0669359$ (Fig.~\ref{fig:curv}). Figure~\ref{fig:tors} shows $\tau_1$ approaching these
limits and Fig.~\ref{fig:dom} the timelike domain. The $a/M=3/2$ spacetime has no event horizon, so for these curves the curvature divergence with finite torsion is a feature of nullness and not of a horizon.
Two covariant constructions (Eq.~\eqref{eq:tau1def} and Eq.~\eqref{eq:tau1B}) give identical rational expressions, and an independent high-precision numerical implementation reproduces Eq.~\eqref{eq:rational}
to twelve digits at five radii between $r=1.62$ and $2.5$; Eq.~\eqref{eq:tauStar} reproduces both limits to nine digits.
Double-precision finite differencing of $De_1/d\tau$ near the null point is ill-conditioned, so all results here use at least 25-digit arithmetic and are stable at 25, 40 and 60 digits.


\section{Horizon crossing and the extremal limit}
\label{sec:horizon}

A horizon-penetrating chart removes the coordinate singularity from the metric components at the future horizon; it does not by itself guarantee finite FS invariants for every curve. The precise local statement is:
if a worldline is smooth through the horizon in a regular chart, remains timelike there, and has $\kappa>0$ with a nondegenerate FS frame, then $\kappa$ and $\tau_1$ are finite at the crossing, being smooth functions of the metric up to its second derivatives and the
curve's derivatives up to third order.
For the KS-form spiral with $(a,\omega,\lambda)=(0.5,0.1,-0.45)$ the norm is positive at $r_+=1.8660254$, where $\kappa M=0.97235$ and $\tau_1M=0.19411$, and the invariants vary smoothly
across $r_+$ until the curve becomes null at $r=1.6062$ (Fig.~\ref{fig:two}). The extremal case was treated in Sec.~\ref{sec:killing}: the fixed-$\Omega_H$ tangent is spacelike outside the horizon, so a valid extremal limit requires a timelike family, and the limits $a\to M$ and
$r\to r_+$ need not commute.

\section{Gyroscope observables}
\label{sec:obs}

The FS quantities describe a chosen worldline; by themselves, they are not detector signals. Here we explain how they are related to the rotation of a gyroscope relative to specified reference directions.

\subsection{What \texorpdfstring{$\tau_1$}{tau1} measures}

A gyroscope spin $S^\mu\perp u$ obeys Fermi--Walker transport, $DS^\mu/d\tau=-(A_\nu S^\nu)u^\mu$. The FS vectors obey $D_{\rm FW}e_1\equiv De_1/d\tau-\kappa u=\tau_1e_2$, so in the equatorial case the gyroscope axis rotates at rate $\tau_1$ relative to $e_1$,
the direction of the proper acceleration, i.e.\ the axis that an accelerometer on board measures. $\tau_1$ is therefore the rotation rate of the gyroscope relative to the accelerometer axis. Where $\kappa\to0$, $e_1$ is ill-defined
and $\tau_1$ is not a property of the gyroscope's rotation (Sec.~\ref{sec:spike}).

\subsection{A regular geometric frame}

The Kerr geometry singles out the principal null directions. The ingoing one is, in BL components, $((r^2+a^2)/\Delta,-1,0,a/\Delta)$; with Eq.~\eqref{eq:uKS} it becomes $l=(1,-1,0,0)$ in the ingoing chart, which is regular on the future horizon. For any timelike $u$
the projection $w_1=[l-(l\!\cdot\! u)u]/|l\!\cdot\! u|$ is a unit vector orthogonal to $u$ and regular wherever the curve is timelike. Completing $(u,w_1)$ to an orthonormal triad $(u,w_1,w_2)$ in the equatorial 3-space
defines a regular frame. The gyroscope rotates relative to $w_1$ at the rate $\beta\equiv g(w_2,Dw_1/d\tau)$. With $\vartheta$ the angle of $e_1$ relative to $w_1$ we find
\begin{equation}
\beta=\dot\vartheta+\tau_1 ,\label{eq:beta}
\end{equation}
confirmed to eight digits at $r=3.0,\,2.0,\,1.8$ for the KS-form spiral $(0.5,0.1,-0.45)$ (e.g.\ at $r=1.8$, inside the horizon: $\beta=0.099259$, $\dot\vartheta=-0.153718$, $\tau_1=0.252976$). Thus the gyroscope's rotation relative to a
geometric frame equals its rotation relative to the accelerometer axis plus the rotation of that axis relative to the frame; all three are finite across the horizon.

\subsection{Geodesic plunge and Marck's frame}

\begin{figure}[t]
\includegraphics[width=\columnwidth]{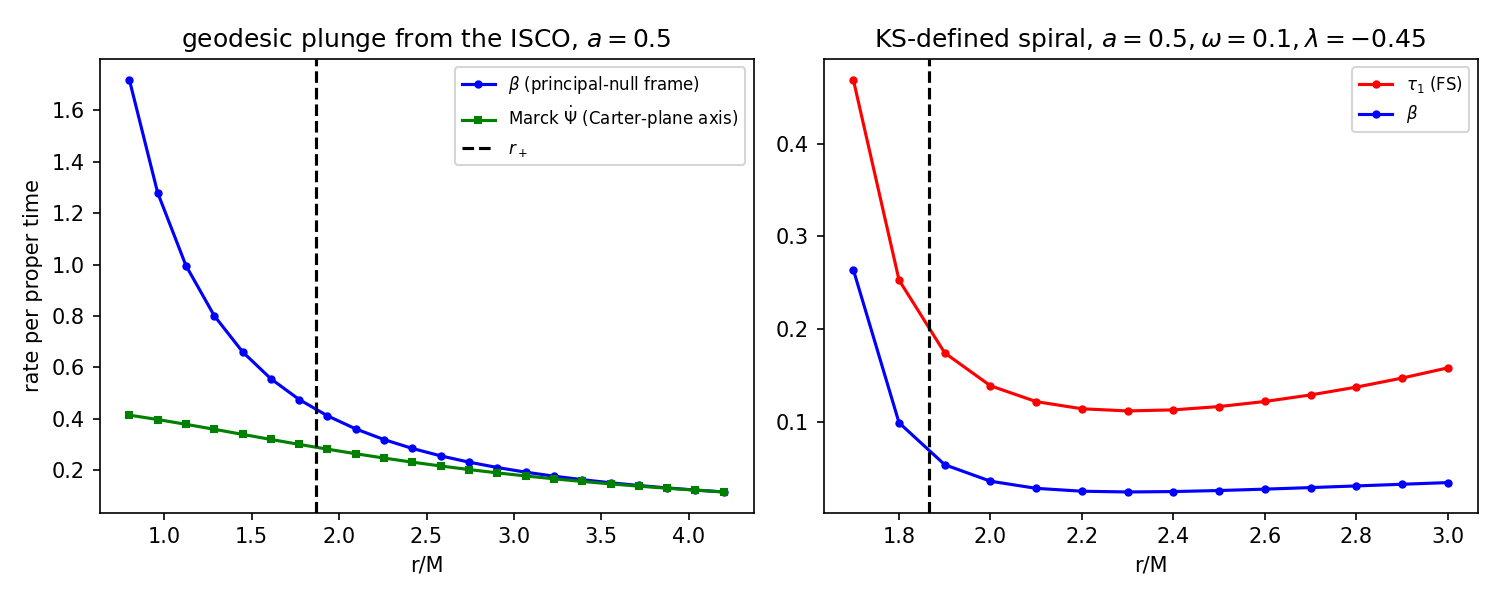}
\caption{Left: geodesic plunge from the ISCO ($a=0.5$): $\beta$ (principal-null frame) and Marck's $\dot\Psi$, both finite across $r_+$ (dashed). Right: KS-form spiral $(0.5,0.1,-0.45)$: $\tau_1$ and $\beta$.}
\label{fig:rates}
\end{figure}

For a geodesic $\kappa=0$, so FS torsion is undefined, but $\beta$ is well defined. We consider the equatorial plunge with $E,L$ of the prograde ISCO for $a=0.5$
($r_{\rm isco}=4.23300253$, $E=0.9178820067$, $L=2.902866153$), with components transformed to the ingoing chart; $\kappa^2<10^{-40}$ along it. Marck's frame is built from the Killing--Yano tensor~\cite{Marck,Bini17}; in the equatorial plane its in-plane axis is the unit vector in the
Carter $t$--$r$ plane orthogonal to $u$, i.e.\ in $\mathrm{span}\{l_{\rm in},l_{\rm out}'\}$ with $l_{\rm out}'=\Delta\,l_{\rm out}\propto(r^2+a^2+2Mr,\Delta,2a)$ in the ingoing chart (nonzero on the horizon). We find that the rotation rate of a gyroscope relative to this axis
equals Marck's analytic rate
\begin{equation}
\dot\Psi=\frac{\sqrt K}{r^2}\Bigl[\frac{E(r^2+a^2)-aL}{K+r^2}+\frac{a(L-aE)}{K}\Bigr],\quad K=(L-aE)^2,\label{eq:Marck}
\end{equation}
to ten digits at every radius tested, including at and inside the horizon (Table~\ref{tab:plunge}); this supports identifying the Carter-plane axis with the in-plane axis of Marck's frame \cite{Marck}.
Both $\beta$ and $\dot\Psi$ are finite and smooth through $r_+$ (Fig.~\ref{fig:rates}); they differ because the two reference axes rotate relative to each other. Over $r=3\to1$ (proper time $5.95$) the accumulated angles are
$\int\beta\,\dd\tau\simeq1.89$ and $\int\dot\Psi\,\dd\tau\simeq1.39$ rad; along the KS-form spiral over $r=3\to1.7$ (proper time $6.45$), $\int\tau_1\dd\tau\simeq0.94$ and $\int\beta\,\dd\tau\simeq0.26$ rad (low-order quadrature, two-digit accuracy).
Angles are meaningful only relative to a stated reference direction, and these are not detector outputs.

\begin{table}[t]
\caption{Equatorial plunge from the ISCO, $a=0.5$: rotation rate relative to the principal-null frame ($\beta$) and Marck's rate $\dot\Psi$, Eq.~\eqref{eq:Marck}, which equals the rate
relative to the Carter-plane axis. $r_+=1.8660254$.}
\label{tab:plunge}
\begin{ruledtabular}
\begin{tabular}{ccc}
$r$ & $\beta$ & $\dot\Psi$\\
\hline
3.0 & 0.2002828 & 0.1832149\\
2.2 & 0.3333841 & 0.2537032\\
$r_+$ & 0.4366892 & 0.2901414\\
1.5 & 0.6236979 & 0.3336144\\
1.0 & 1.2019987 & 0.3934211\\
\end{tabular}
\end{ruledtabular}
\end{table}

\subsection{Near-geodesic points}
\label{sec:spike}

$\tau_1^2$ contains $1/\kappa^2$. For the KS-form spiral $(a,\omega,\lambda)=(2.5,0.3,-0.15)$ the $\tau_1$ spike at $r\simeq0.783$ coincides with a minimum of $\kappa$ ($\kappa=0.0873$; $\tau_1=5.50$ there, $\kappa\tau_1\simeq0.48$), a near-geodesic point at which the FS frame spins up.
The rotation rate relative to the regular frame is smooth there ($\beta\simeq1.345$ at $r=0.78$). The spike is therefore a property of the FS frame, not of the gyroscope. This is relevant to EMRIs: the orbit is geodesic to leading order
(an ISCO plunge is exactly geodesic), so FS torsion is undefined or ill-conditioned, and the radiation-reaction acceleration, being small, would set the direction of $e_1$. We expect regular-frame rates of the type $\beta$ to be the appropriate quantities there; we have not computed a flux-driven inspiral.

\section{Discussion and conclusions}
\label{sec:disc}

This study has led to five main results. First, Eq.~\eqref{eq:kappaS} gives an exact expression for the curvature in terms of a non-unit tangent. Second, when $S'\ne0$ at a null endpoint, Eq.~\eqref{eq:tauStar} gives a finite limit for the first torsion; we checked it against two exact symbolic examples and seven additional cases. Third, the invariants agree under coordinate transformations in five charts, including a $\theta$-dependent chart, and in an off-equatorial test. Fourth, the closed expressions for Killing observers show that both $\kappa$ and $\tau_1$ diverge at light surfaces, which can also occur when $a>M$. Finally, we relate $\tau_1$ to rotation relative to the acceleration direction and compare it with a rate defined using a regular geometric frame. That latter rate remains finite through the horizon on a geodesic plunge and agrees with Marck's result.

For the examples considered here, the key issue is whether the prescribed worldline approaches a null curve, not simply whether a horizon is present. In both the sub-extremal and super-extremal spiral examples, $\kappa$ diverges while $\tau_1$ tends to a finite value. The BL curves that appear to develop a divergence near the horizon become null before reaching it; they are not coordinate descriptions of the same curves used in the horizon-penetrating chart. These results do not, on their own, provide an observable that distinguishes a black hole from a horizonless compact object.

\emph{Limitations.} The analysis is equatorial except for one off-equatorial chart-independence and null-limit test. Proposition~\ref{prop:tau} is derived by a symbolic series expansion, with numerical validation but without a bound on higher-order terms. The Killing-horizon law is verified numerically
(Kerr and Kerr--Newman) rather than proved. The extremal case needs a separate treatment. A numerical double-null comparison is not attempted. The spiral and plunge trajectories are kinematic or geodesic, not self-force driven; no waveform or detector observable is computed.
The next steps are to extend the analysis to general non-equatorial trajectories, replace the prescribed spirals with an inspiral driven by radiation-reaction forces, and implement the double-null coordinate comparison.

\begin{acknowledgments}
The author thanks K.\ Rajesh Nayak for many useful discussions and the Centre of Excellence in Space Sciences India (CESSI), IISER Kolkata.
\end{acknowledgments}

\textbf{Data availability.} The scripts for the symbolic and high-precision numerical calculations, with their outputs, will be given upon request.

\appendix

\section{Derivation of Proposition~\ref{prop:tau}}
\label{app:der}

Write $S'=2Q$, $p=D_vq$ and use the identities $q\!\cdot\! v=Q$, $v\!\cdot\! p=Q'-q\!\cdot\! q$ and $q\!\cdot\! p=\tfrac12(q\!\cdot\! q)'$. With $u=v/\sqrt S$ and $\dd\tau=\sqrt S\,\dd s$,
\begin{equation}
A=\frac{q-(Q/S)v}{S},\qquad
\kappa^2=\frac{Q^2}{S^3}-\frac{q\!\cdot\! q}{S^2}.
\end{equation}
Differentiating once more,
\begin{equation}
B=\frac{DA}{d\tau}=\frac{W}{S^{3/2}},\quad W=p-\frac{3Q}{S}\,q-C\,v,\quad C=\frac{Q'}{S}-\frac{4Q^2}{S^2},
\end{equation}
so that $B\!\cdot\! B=W\!\cdot\! W/S^3$ with
\begin{align}
W\!\cdot\! W&=p\!\cdot\! p+\frac{9Q^2\,q\!\cdot\! q}{S^2}+C^2S-\frac{3Q}{S}(q\!\cdot\! q)'\nonumber\\
&\quad-2C\,(Q'-q\!\cdot\! q)+\frac{6CQ^2}{S}.
\end{align}
Finally $\dot\kappa^2=(\dd\kappa/\dd s)^2/S=(\dd\kappa^2/\dd s)^2/(4\kappa^2S)$ and $\tau_1^2=(\kappa^4-\dot\kappa^2-B\!\cdot\! B)/\kappa^2$. All quantities are functions of $S(s)$, $q\!\cdot\! q(s)$ and $p\!\cdot\! p(s)$ only.
Expanding $S=n_1\sigma+\tfrac12n_2\sigma^2+\dots$, $q\!\cdot\! q=B_0+B_1\sigma+\dots$ and $p\!\cdot\! p=D_0+\dots$ in $\sigma=s-s_0$ (computer algebra, SymPy) the terms of order $\sigma^{-6},\sigma^{-5},\sigma^{-4}$ in
$\kappa^4-\dot\kappa^2-B\!\cdot\! B$ cancel identically and the $\sigma^{-3}$ coefficient divided by $\kappa^2\simeq1/(4n_1\sigma^3)$ yields Eq.~\eqref{eq:tauStar}; no coefficient beyond $n_2$, $B_0$, $B_1$, $D_0$ survives. (Here $B_1=\dd(q\!\cdot\! q)/\dd s|_{s_0}=2q\!\cdot\! p$.)

\section{Polynomials for the two spirals}
\label{app:poly}

Writing $P(r)=\sum_{k=0}^{11}p_kr^k$ and $R(r)=\sum_{k=0}^{11}\rho_kr^k$, the exact results are
$\tau_1=\tfrac12P_{1/2}/(r^2R_{1/2})$ for $a/M=1/2$ (Eq.~\eqref{eq:rational}) and $\tau_1=\tfrac32P_{3/2}/(r^2R_{3/2})$ for $a/M=3/2$, with the primitive integer coefficients below.
They are generated, and the identities checked symbolically, by the script \texttt{sympy\_exact\_rational.py} in the repository.

\begin{table*}[h]
\caption{Integer coefficients of $P$ and $R$ for $(a/M,\omega M,\lambda)=(1/2,\,1/10,\,-9/20)$ and $(3/2,\,3/10,\,-9/20)$.}
\scriptsize
\begin{ruledtabular}
\begin{tabular}{rrrrr}
$k$ & $p_k$ ($a/M=1/2$) & $\rho_k$ ($a/M=1/2$) & $p_k$ ($a/M=3/2$) & $\rho_k$ ($a/M=3/2$)\\
\hline
11 & 8069871680 & 370177600 & 761603394240 & 269859470400\\
10 & $-26772037120$ & 49254400 & $-149315736960$ & 641827929600\\
9 & 174273128 & $-43184511776$ & $-645885790488$ & $-2412570723264$\\
8 & $-156465270888$ & 48627189168 & $-706497559272$ & $-1212884672688$\\
7 & 1833728643912 & $-30703052415$ & 5666030385768 & $-4110527216223$\\
6 & $-3433636613679$ & 4890095876382 & $-15272208162639$ & 16683618816342\\
5 & 62941115872140 & $-10003404494220$ & 30508038200420 & $-28701078945660$\\
4 & $-244350791458500$ & 13645866063000 & $-36626587046500$ & 71045818767000\\
3 & 214616875380000 & $-352080083490000$ & 15119973660000 & $-137816305330000$\\
2 & 267021212950000 & 1019540279300000 & 24973885750000 & 169772643700000\\
1 & $-389737982600000$ & $-831474044200000$ & $-37653723800000$ & $-117341758600000$\\
0 & 47045881000000 & 94091762000000 & 15944049000000 & 31888098000000\\
\end{tabular}
\end{ruledtabular}
\end{table*}

\section{Numerical methods and checks}
\label{app:num}

All curves in this paper have $r$- (and, in one test, $\theta$-) dependent tangent fields. For an equatorial curve, $\theta$-derivatives of the metric vanish and the problem reduces to the $(t,r,\phi)$ block; Christoffel symbols are obtained analytically from the pulled-back metric, and
derivatives along the curve ($u^r\partial_r$, plus $u^\theta\partial_\theta$ off the equator) are computed by high-order numerical differentiation at 30--50 significant digits. $\tau_1$ is evaluated from Eq.~\eqref{eq:tau1def}. Checks performed:
(i) at $\lambda\to0$ the spiral reproduces the Killing closed forms, and Eq.~\eqref{eq:kapK} reproduces the $\kappa^2$ of Eq.~(12) of Ref.~\cite{prev};
(ii) for $a=0$, $\omega=\lambda=0$, $\kappa^2=M^2/[r^4(1-2M/r)]$ and $\tau_1=0$;
(iii) the pulled-back metric agrees with the Kerr--Schild metric of Ref.~\cite{prev} (Eq.~(23) there) to $10^{-19}$ and its normalisation $V^2$ (Eq.~(25) there) agrees with the numerical norm;
(iv) Eq.~\eqref{eq:rational} against the numerical $\tau_1^2$ at $r=2.5,2.0,1.8,1.7,1.62$ (twelve digits);
(v) $\beta=\dot\vartheta+\tau_1$, Eq.~\eqref{eq:beta}, and Eq.~\eqref{eq:Marck} against the numerical rates (eight and ten digits).
The closed forms \eqref{eq:kapK}--\eqref{eq:diffK} were obtained symbolically and checked against the numerical FS code.

\end{document}